\documentclass[twocolumn,preprintnumbers,amsmath,amssymb,pra,floatfix]{revtex4}

\usepackage{graphicx}
\usepackage{dcolumn}
\usepackage{bm}

\begin{document}

\title{The Magnetic Casimir Effect}  

\author{G. Metalidis}
\author{P. Bruno}
  \email{bruno@mpi-halle.de}
\affiliation{%
Max-Planck-Institut f\"{u}r Mikrostrukturphysik, Weinberg 2,
D-06120 Halle, Germany}
\homepage{http://www.mpi-halle.de}
\date{\today}

\begin{abstract}
The Casimir effect results from alterations of the zero-point
electromagnetic energy introduced by boundary-conditions. For
ferromagnetic layers separated by vacuum (or a dielectric) such
boundary-conditions are influenced by the magneto-optical Kerr
effect. We will show that this gives rise to a long-range magnetic
interaction and discuss the effect for two different
configurations (magnetization parallel and perpendicular to the
layers). Analytical expressions are derived for two models and
compared to numerical calculations. Numerical calculations of the
effect for Fe are also presented and the possibility of an
experimental observation of the Casimir magnetic interaction is
discussed.
\end{abstract}

\maketitle

\section{Introduction}
Since its discovery, the Casimir effect has gradually become a
much-discussed subject in physics. Originally, one understood by
the Casimir effect the attractive force between two metal plates
in vacuum as a result of zero-point quantum
fluctuations~\cite{HBGCas}. Nowadays the term is used for a much
broader range of effects, all involving the influence of
boundaries on fluctuations. As such, the Casimir effect plays a
role in quantum field theory, atomic and molecular physics,
condensed matter physics, gravitation, cosmology and so on. A
thorough review of the Casimir effect in all these fields was
published recently~\cite{MBor}.

For two uniformly magnetized ferromagnetic plates held parallel to
each other, it is shown in a previous paper~\cite{PBru} that the
interplay of the Casimir effect and the magneto-optical Kerr
effect gives rise to a new long-range magnetic interaction. In
[\onlinecite{PBru}], this magnetic Casimir force was found to
decay with interplate distance D as $D^{-5}$ in the limit of long
distances, and as $D^{-1}$ for short distances. In this case, the
ferromagnetic plates were described with a Drude model and the
magnetization was defined to be perpendicular to the plates. In
view of future experimental investigations of this new magnetic
Casimir force, it would be useful to study the case where the
magnetization is parallel to the plates since this situation is
easier to obtain in an experimental setup. This subject will be
studied in the present paper and a force which decays as $D^{-6}$
in the long-distance limit and as $D^{-3}$ in the limit of short
distances is found when the Drude model is used. This behavior is
interesting since it means that the force is larger, and thus
easier to measure, for in-plane magnetization than for
perpendicular magnetization at sufficiently small distances. Next
to the Drude model, another more realistic model is also studied.
In this so called hybrid model, a plasma model is used for the
diagonal element of the dielectric tensor of the magnetic plates,
and a single absorption line model for the off-diagonal element.
As for the long distance limit, the force in this model goes like
$D^{-8}$ for the case of perpendicular magnetization and as
$D^{-10}$ when the magnetization is parallel to the plates while
the behavior in the short-distance limit is unchanged. Finally we
will present some numerical calculations of the interaction for
Fe, in which experimental data for the elements of the dielectric
tensor are used. An experimental setup to measure the magnetic
Casimir force is also discussed.

\section{General Theory}
Consider two uniformly magnetized ferromagnetic plates of infinite
lateral extension held parallel to each other. The Casimir
interaction energy per unit area at $T=0$ between the plates can
be expressed as~\cite{MTJae}:
\begin{widetext}
\begin{equation}
{\cal E} = \frac{\hbar}{(2\pi)^{3}}\int_{0}^{+\infty}{\rm d}\omega
\int\, {\rm d}^{2}\bm{k}_{\parallel}\,\mbox{Im Tr} \ln(1-{\sf
R}_{A} {\sf R}_{B}{\rm e}^{2{\rm i} k_{\perp}D}) , \label{eq1}
\end{equation}
\end{widetext}
where $k_{\perp}$ and $\bm{k}_{\parallel}$ are the components of
the wavevector perpendicular and parallel to the mirrors. The
$2\times2$ matrices ${\sf R}_{A}$ and ${\sf R}_{B}$ contain the
reflection coefficients of the two mirrors:
\begin{equation}
{\sf R}_{A(B)}=\left(\begin{array}{cc}
                r_{ss}^{A(B)} & r_{sp}^{A(B)} \\
                r_{ps}^{A(B)} & r_{pp}^{A(B)}
              \end{array}\right)   \label{eq2}
\end{equation}
The index $s$ (resp. $p$) corresponds to a polarization with the
electric field perpendicular (resp. parallel) to the incidence
plane. We will adopt here the usual convention that the $s$ axis
remains unchanged upon reflection. Since the reflection
coefficients are dependent on the direction of the magnetisation
of the mirrors, it is clear from Eqs.~(\ref{eq1}) and ~(\ref{eq2})
that the magnetic Casimir energy between the mirrors will differ
for the situations in which the magnetizations of the two mirrors
are parallel (FM) or antiparallel (AF). This will result in a net
magnetic Casimir force per unit area $\Delta{\cal F} \equiv {\cal
F}_{AF} - {\cal F}_{FM}$ between the mirrors, different from the
ordinary Casimir force discussed in Ref.~[\onlinecite{HBGCas}].
\\If a change of integration variables
$(\omega,\bm{k}_{\parallel})\longrightarrow(\omega,k_{\perp},\varphi)$
is performed and complex integration methods are used as
in~[\onlinecite{MTJae}], Eq.~(\ref{eq1}) can be written as:
\begin{widetext}
\begin{equation}
{\cal E}=\frac{\hbar}{(2\pi)^{3}} \int_{0}^{+\infty} {\rm d}
k_{\perp}\,k_{\perp}  \int_{0}^{2\pi} {\rm d}\varphi
\int_{0}^{k_{\perp}c} {\rm d}\omega\, \mbox{Re Tr} \ln[1-{\sf
R}_{A}({\rm i}\omega,{\rm i}k_{\perp},\varphi) {\sf R}_{B}({\rm
i}\omega,{\rm i}k_{\perp}, \varphi){\rm e}^{-2k_{\perp}D}]
\label{eq3}
\end{equation}
\end{widetext}
In general, the reflection coefficients contain terms of different
orders of the magneto-optical constant Q. In our calculation, only
terms up to first order in Q will be conserved. When the
magnetisation direction is reversed, these terms will change sign.
Since the first order terms are usually much smaller than 1 and
than the terms that are independent of Q, it is possible to expand
expression~(\ref{eq3}) to lowest order in the linear terms.

\subsection{The Polar Configuration}
After some algebra, we find for the situation with magnetisation
directed perpendicular to the plates (we will call this the polar
configuration from now on):
\begin{widetext}
\begin{subequations}
\label{eq4}
\begin{eqnarray}
\Delta{\cal E}^{\perp} =& {\cal E}_{AF}^{\perp}-{\cal
E}_{AM}^{\perp} &\approx
-\frac{\hbar}{\pi^{2}}\int_{0}^{+\infty}{\rm d}k_{\perp}k_{\perp}
\,\int_{0}^{k_{\perp}c}{\rm d}\omega\, \mbox{Re}
\left[\frac{(r_{sp}^{\perp})^{2}{\rm e}^{-2k_{\perp}D}}
{(1-r_{ss}^{2} {\rm e}^{-2k_{\perp}D}) (1-r_{pp}^{2} {\rm
e}^{-2k_{\perp}D})}
\right] ,  \label{eq4a} \\
\Delta{\cal F}^{\perp} =&  -\frac{{\rm d}\Delta{\cal
E}^{\perp}}{{\rm d}D} &\approx
-\frac{2\hbar}{\pi^{2}}\int_{0}^{+\infty}{\rm
d}k_{\perp}\,k_{\perp}^{2} \int_{0}^{k_{\perp}c}{\rm
d}\omega\,\mbox{Re}\left[\frac{(r_{sp}^{\perp})^{2} [1-r_{ss}^{2}
r_{pp}^{2}{\rm e}^{-4k_{\perp}D}]{\rm e}^{-2k_{\perp}D}}
{([1-r_{ss}^{2}{\rm e}^{-2k_{\perp}D}] [1-r_{pp}^{2} {\rm
e}^{-2k_{\perp}D}])^{2}}\right], \label{eq4b}
\end{eqnarray}
\end{subequations}
where the reflection coefficients have to be evaluated at
imaginary perpendicular wavevector and frequency. In this
equation, the reflection amplitudes are supposed to be identical
for the two mirrors. Otherwise, the squared reflection
coefficients have to be replaced by the product of the
coefficients for the separate mirrors (e.g. $r_{ss}^{2}\rightarrow
r_{ss}^{A}r_{ss}^{B}$). The integral over the angle $\varphi$ is
already performed. The reflection coefficients for a mirror in the
polar configuration are given in~[\onlinecite{GMet}] as:
\begin{subequations}
\label{eq5}
\begin{eqnarray}
r_{ss}({\rm i}\omega,{\rm i}k_{\perp}) &=&
\frac{k_{\perp}c-\xi}{k_{\perp} c+\xi}
,\ \ \ \ \ \ 
r_{pp}({\rm i}\omega,{\rm i}k_{\perp}) =
\frac{\varepsilon_{xx}({\rm
i}\omega)k_{\perp}c-\xi}{\varepsilon_{xx}({\rm
i}\omega)k_{\perp}c+\xi} \label{eq5a}\\
r_{sp}^{\perp}({\rm i}\omega,{\rm i}k_{\perp})&=&
r_{ps}^{\perp}({\rm i}\omega,{\rm i} k_{\perp})= \frac{-k_{\perp}
c\,\omega\,\varepsilon_{xy}({\rm
i}\omega)}{\left[k_{\perp}c+\xi\right]\left[\varepsilon_{xx} ({\rm
i}\omega)k_{\perp}c+\xi\right]}, \label{eq5b}
\end{eqnarray}
\end{subequations}
with: $\xi=\sqrt{\omega^{2} (\varepsilon_{xx}({\rm
i}\omega)-1)+(k_{\perp}c)^{2}}$
\end{widetext}

\subsection{The In-plane Configuration}
 For the case where the magnetisation is parallel to the plates
(we will refer to this situation as the in-plane configuration
from now on), not only $r_{sp}$, but also $r_{pp}$ will contain a
term that is linear in the magneto optical constant. As a
consequence we find two contributions to the Casimir magnetic
interaction energy. The first one ($\Delta{\cal
E}_{1}^{\parallel}$) results from the longitudinal Kerr effect,
while the second term ($\Delta{\cal E}_{2}^{\parallel}$) is a
consequence of the transversal Kerr effect. Again the integral
over $\varphi$ can be performed directly. We obtain (for identical
mirrors):
\begin{widetext}
\begin{subequations}
\label{eq6}
\begin{eqnarray}
\Delta{\cal E}_{1}^{\parallel} &\approx&
\frac{\hbar}{2\pi^{2}}\int_{0}^{+\infty}{\rm
d}k_{\perp}k_{\perp}\, \int_{0}^{k_{\perp}c}{\rm d}\omega\,
\mbox{Re}\left[\frac{(r_{sp}^{\parallel})^{2} {\rm
e}^{-2k_{\perp}D}} {(1-r_{ss}^{2}{\rm e}^{-2k_{\perp}D})
(1-r_{pp}^{2}{\rm e}^{-2k_{\perp}D})}\right] ,
\label{eq6a} \\
\Delta{\cal E}_{2}^{\parallel} &\approx&
\frac{-\hbar}{4\pi^{2}}\int_{0}^{+\infty}{\rm d}k_{\perp}k_{\perp}
\, \int_{0}^{k_{\perp}c}{\rm d}\omega\,Re\left[ \frac {\Delta
r_{pp}^{2} {\rm e}^{-2k_{\perp}D}} {(1-r_{pp}^{2} {\rm
e}^{-2k_{\perp}D})^{2}} \right]
,\label{eq6b} \\
\Delta{\cal F}_{1}^{\parallel} &\approx& \frac{\hbar}{\ \pi^{2}}
\int_{0}^{+\infty}{\rm d}k_{\perp}\,k_{\perp}^{2}\,
\int_{0}^{k_{\perp}c}{\rm d}\omega\,
\mbox{Re}\left[\frac{(r_{sp}^{\parallel})^{2}[1-r_{ss}^{2}
r_{pp}^{2}{\rm e}^{ -4k_{\perp}D}] {\rm e}^{
-2k_{\perp}D}}{([1-r_{ss}^{2}{\rm e}^{ -2k_{\perp}D}]
[1-r_{pp}^{2}{\rm e}^{-2k_{\perp}D}])^{2}}\right]
, \label{eq6c} \\
\Delta{\cal F}_{2}^{\parallel} &\approx&
\frac{-\hbar}{2\pi^{2}}\int_{0}^{+\infty}{\rm d}
k_{\perp}\,k_{\perp}^{2} \int_{0}^{k_{\perp}c}{\rm d}\omega\,
\mbox{Re}\left[\frac{\Delta r_{pp}^{2} (1+r_{pp}^{2}{\rm
e}^{-2k_{\perp}D}) {\rm e}^{ -2k_{\perp}D}}{(1-r_{pp}^{2} {\rm e}^
{-2k_{\perp}D})^{3}}\right]. \label{eq6d}
\end{eqnarray}
\end{subequations}
\end{widetext}
Of course $\Delta{\cal E}^{\parallel}=\Delta{\cal
E}_{1}^{\parallel}+\Delta{\cal E}_{2}^{\parallel}$ and
$\Delta{\cal F}^{\parallel}=\Delta{\cal
F}_{1}^{\parallel}+\Delta{\cal F}_{2}^{\parallel}$. For two
different mirrors, the squares of the reflection coefficients have
to be replaced as mentioned above. The integration over the angle
$\varphi$ is already performed. The reflection coefficients in
Eq.~(\ref{eq6}) are given in~[\onlinecite{GMet}]:
\begin{widetext}
\begin{subequations}
\label{eq7}
\begin{eqnarray}
r_{sp}^{\parallel}({\rm i}\omega,{\rm i}k_{\perp}) &=&
-r_{ps}^{\parallel}({\rm i}\omega,{\rm i}k_{\perp}) = (-1)\frac{
\sqrt{\omega^{2}-(k_{\perp}c)^{2}} \, \omega\,
\varepsilon_{xy}({\rm i}\omega)\,(k_{\perp}c)}
{\left(k_{\perp}c+\xi\right)\left(\varepsilon_{xx} ({\rm
i}\omega)k_{\perp}c+\xi\right)\xi}, \label{eq7a} \\
\Delta r_{pp}({\rm i}\omega,{\rm i}k_{\perp}) &=&
\frac{2\sqrt{\omega^{2}-(k_{\perp}c)^{2}} \, \varepsilon_{xy}({\rm
i}\omega) \, (k_{\perp}c)} {\left(\varepsilon_{xx}
(i\omega)k_{\perp}c+\xi\right)^{2}}, \label{eq7b}
\end{eqnarray}
\end{subequations}
\end{widetext}
with again: $\xi=\sqrt{\omega^{2} (\varepsilon_{xx}({\rm
i}\omega)-1)+(k_{\perp}c)^{2}}$. $r_{ss}(i\omega,ik_{\perp})$ and
$r_{pp}(i\omega,ik_{\perp})$ are still given by Eq.~(\ref{eq5a}).
Note that the contributions arising from the longitudinal and
transversal Kerr effect are of opposite sign and therefore tend to
cancel each other. On the basis of Eqs.~(\ref{eq4}-\ref{eq7}), we
will calculate the Casimir magnetic energies and forces for two
simple models in the next two sections. In
section~\ref{sectionnumeric}, we will use these equations to
numerically calculate the interaction for iron plates.

\section{The Drude Model}
Consider two identical magnetic mirrors with a dielectric tensor
described by the Drude model:
\begin{subequations}
\label{eq8}
\begin{eqnarray}
\varepsilon_{xx}({\rm i}\omega) &=& 1 +
\frac{\omega_{p}^{2}\tau}{\omega(1+\omega\tau)}, \label{eq8a} \\
\varepsilon_{xy}({\rm i}\omega) &=&
\frac{\omega_{p}^{2}\omega_{c}\tau^{2}}{\omega(1+\omega\tau)^{2}}.
\label{eq8b}
\end{eqnarray}
\end{subequations}
In this equation $\omega_{p}$ is the plasma frequency defined by
$\omega_{p} \equiv 4 \pi n e^{2}/m^{\star}$; $\omega_{c}$ is the
cyclotron frequency given by $\omega_{c} \equiv e B_{{\rm
eff}}/m^{\star}c$ where $B_{{\rm eff}}$ is the effective magnetic
field experienced by the conduction electrons as a result of
exchange and spin-orbit interactions; $\tau$ is the relaxation
time. In the usual situation: $\omega_{c}\tau \ll 1 \ll
\omega_{p}\tau$.

\begin{figure*}
\hspace*{-0.5cm}
\includegraphics{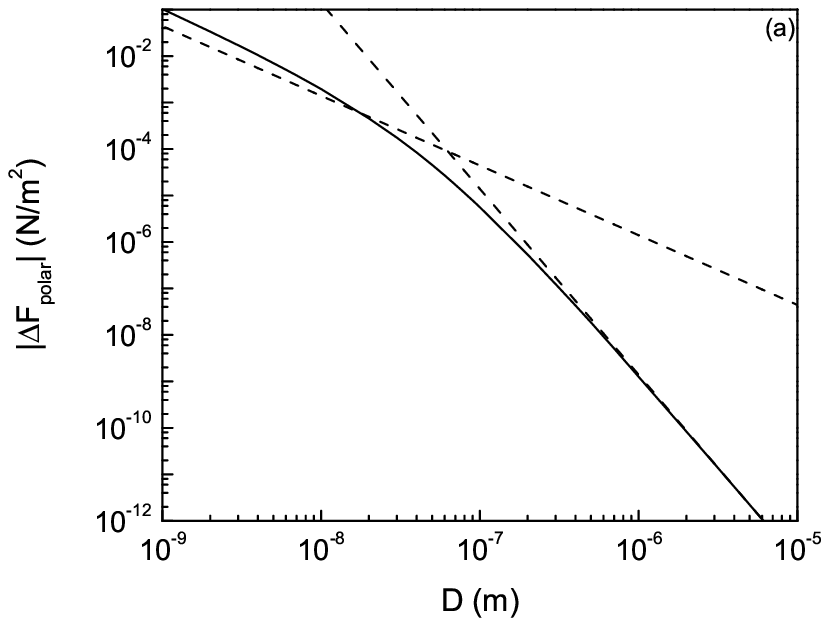}
\hspace*{-1cm}
\includegraphics{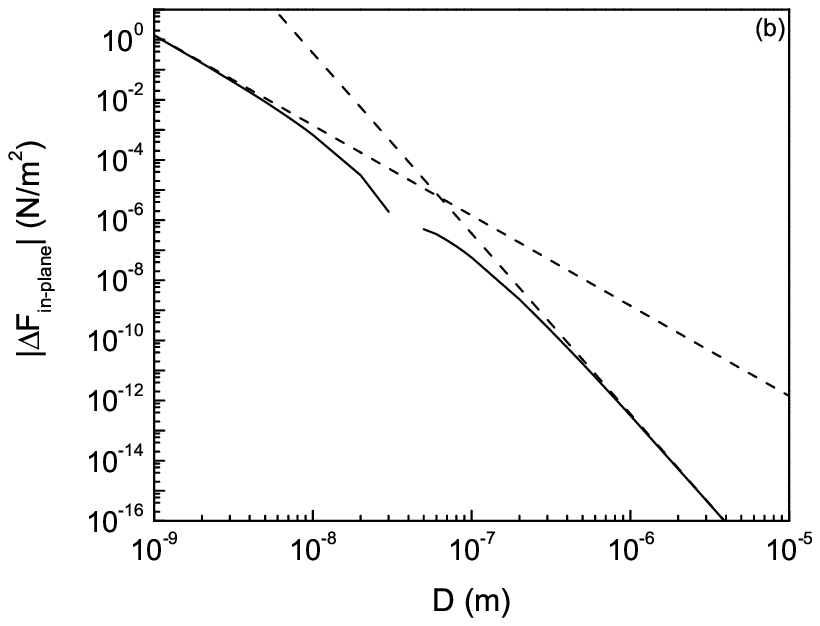}
\vspace*{-\baselineskip}

\hspace*{-0.5cm}
\includegraphics{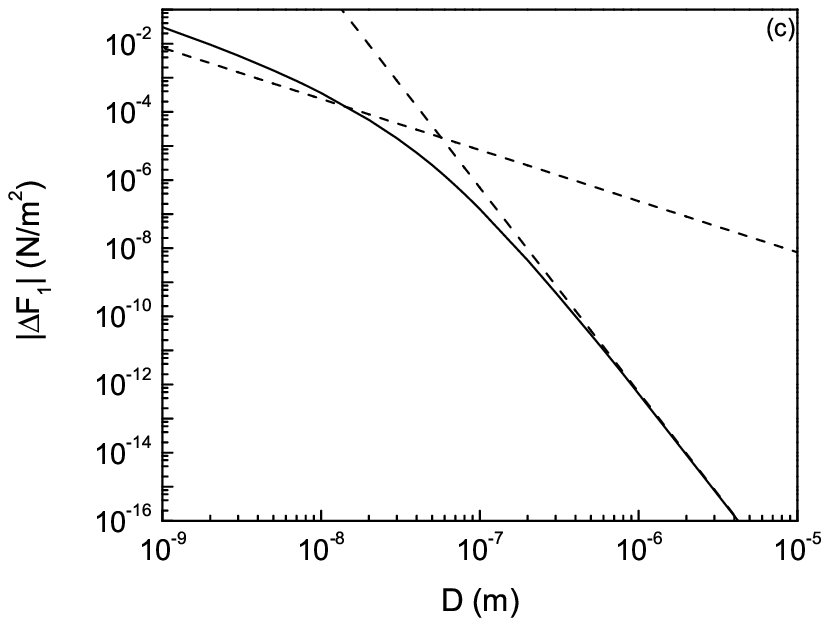}
\hspace*{-1cm}
\includegraphics{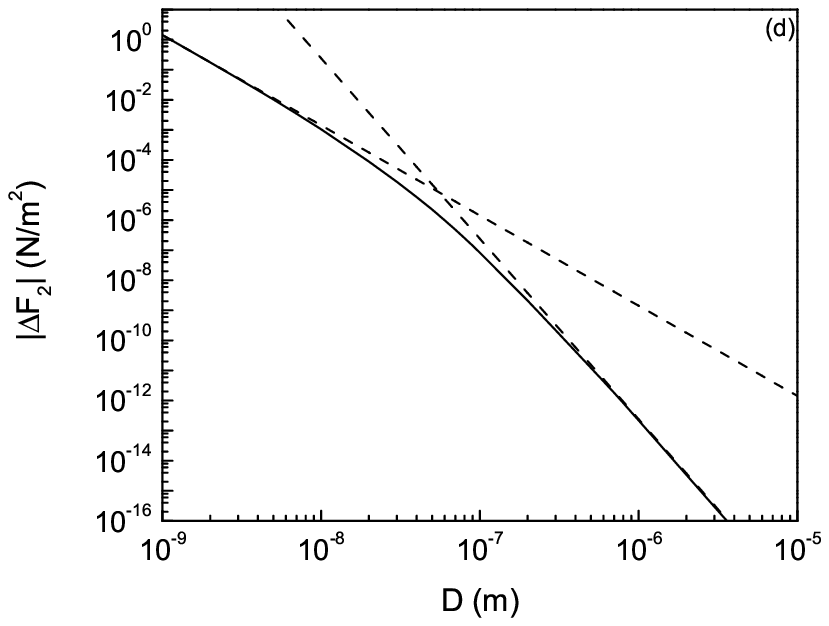}
\vspace*{-\baselineskip}
\caption{\label{fig1}Absolute values of
the magnetic Casimir force (per unit area) with the mirrors
described by a Drude model. Numerical results (solid curves) are
compared with the analytical expressions (dashed curves) for the
polar configuration~(a), the in-plane configuration~(b), the term
$\Delta{\cal F}_{1}^{\parallel}$ resulting from the longitudinal
Kerr effect~(c) and the term from the transversal Kerr effect
$\Delta{\cal F}_{1}^{\parallel}$~(d).}
\end{figure*}

There are three important distance regimes to consider. In the
long distance limit ($D \gg c\tau$) the dominant part in the
integrals in Eqs.~(\ref{eq4}) and~(\ref{eq6}) comes from the
region $\omega \leq k_{\perp} c  \approx c/D \ll 1/\tau$. In this
range, one has:
\begin{subequations}
\label{eq9}
\begin{eqnarray}
\varepsilon_{xx}({\rm i}\omega) &\approx& \varepsilon_{xx}({\rm
i}\omega)-1 \approx \frac{\omega_{p}^{2}\tau}{\omega} \gg 1,
\label{eq9a} \\
\varepsilon_{xy}({\rm i}\omega) &\approx&
\frac{\omega_{p}^{2}\omega_{c}\tau^{2}}{\omega}. \label{eq9b}
\end{eqnarray}
\end{subequations}
With these approximations, one finds for the reflection
coefficients:
\begin{subequations}
\label{eq10}
\begin{eqnarray}
r_{ss}({\rm i}\omega,{\rm i}k_{\perp}) &\approx& -r_{pp}({\rm
i}\omega,{\rm i}k_{\perp}) \approx -1, \label{eq10a} \\
r_{sp}^{\perp}({\rm i}\omega,{\rm i}k_{\perp}) &\approx&
-\frac{\omega_{c}}{\omega_{p}}\sqrt{\omega\tau}, \label{eq10b} \\
r_{sp}^{\parallel}({\rm i}\omega,{\rm i}k_{\perp}) &\approx& -
\frac{\omega_{c}}{\omega_{p}^{2}}\sqrt{\omega^{2} -
(k_{\perp}c)^{2}}, \label{eq10c} \\
\Delta r_{pp}({\rm i}\omega,{\rm i}k_{\perp}) &\approx&
\frac{2\omega_{c}}{\omega_{p}^{2}} \, \frac{\omega
\sqrt{\omega^{2} - (k_{\perp}c)^{2}}}{k_{\perp} c}, \label{eq10d}
\end{eqnarray}
\end{subequations}
For the polar configuration, we arrive at:
\begin{subequations}
\label{eq11}
\begin{eqnarray}
\Delta {\cal E}^{\perp} &\approx& -\frac{3\zeta(3)}{16\pi^2} \,
\frac{{\omega_c}^2\tau}{{\omega_p}^2} \, \frac{\hbar c^2}{D^4}, \label{eq11a}\\
\Delta {\cal F}^{\perp} &\approx& -\frac{3\zeta(3)}{4\pi^2} \,
\frac{{\omega_c}^2\tau}{{\omega_p}^2} \, \frac{\hbar c^2}{D^5} ,
\label{eq11b}
\end{eqnarray}
\end{subequations}
While for the in-plane configuration it is found that:
\begin{subequations}
\label{eq12}
\begin{eqnarray}
\Delta{\cal E}_{1}^{\parallel} &\approx&
-\frac{\zeta(4)}{4\pi^{2}} \frac{\omega_{c}^{2}}{\omega_{p}^{4}}
\, \frac{\hbar c^{3}}{D^{5}}, \ \Delta{\cal E}_{2}^{\parallel}
\approx \frac{\zeta(4)}{10\pi^{2}}
\frac{\omega_{c}^{2}}{\omega_{p}^{4}} \, \frac{\hbar
c^{3}}{D^{5}},
\label{eq12a}\\
\Delta{\cal E}^{\parallel} &=& \Delta{\cal E}_{1}^{\parallel} +
\Delta{\cal E}_{2}^{\parallel} \approx
-\frac{3\zeta(4)}{20\pi^{2}} \,
\frac{\omega_{c}^{2}}{\omega_{p}^{4}} \, \frac{\hbar
c^{3}}{D^{5}},
\label{eq12b}\\
\nonumber\\
\Delta{\cal F}_{1}^{\parallel} &\approx&
-\frac{5\zeta(4)}{4\pi^{2}} \frac{\omega_{c}^{2}}{\omega_{p}^{4}}
\, \frac{\hbar c^{3}}{D^{6}}, \ \Delta{\cal F}_{2}^{\parallel}
\approx \frac{\zeta(4)}{2\pi^{2}}
\frac{\omega_{c}^{2}}{\omega_{p}^{4}} \, \frac{\hbar
c^{3}}{D^{6}}, \ \
\label{eq12c}\\
\Delta{\cal F}^{\parallel} &=& \Delta{\cal F}_{1}^{\parallel} +
\Delta{\cal F}_{2}^{\parallel} \approx -\frac{3\zeta(4)}{4\pi^{2}}
\, \frac{\omega_{c}^{2}}{\omega_{p}^{4}} \, \frac{\hbar
c^{3}}{D^{6}}. \label{eq12d}
\end{eqnarray}
\end{subequations}
The second regime is that for intermediate distances ($c/\omega_p
\ll D \ll c\tau$). Now the integrals in Eqs.~(\ref{eq4})
and~(\ref{eq6}) are dominated by the range $1/\tau \ll\omega \leq
k_\bot c  \approx c/D \ll \omega_p$. For the elements of the
dielectric tensor one then finds:
\begin{subequations}
\label{eq13}
\begin{eqnarray}
\varepsilon_{xx}({\rm i}\omega) &\approx& \varepsilon_{xx}({\rm
i}\omega)-1 \approx \frac{{\omega_p}^2}{\omega^2} \gg 1 , \label{eq13a}\\
\varepsilon_{xy}({\rm i}\omega) &\approx&
\frac{{\omega_p}^2\omega_c}{\omega^3}. \label{eq13b}
\end{eqnarray}
\end{subequations}
In this case, the reflection coefficients $r_{ss}$, $r_{pp}$,
$r_{sp}^{\parallel}$ and $\Delta r_{pp}$ still satisfy
Eqs.~(\ref{eq10a}-\ref{eq10d}), while
\begin{equation}
r_{sp}^{\perp} \approx -\frac{\omega_c}{\omega_p}\label{eq14}.
\end{equation}
Since the reflection coefficients for the in-plane configuration
in this regime are not different from the ones in the short
distance limit, the expressions~(\ref{eq12}) are still valid for
the Casimir magnetic energies and forces in the in-plane
configuration. However, for the polar configuration, one has:
\begin{subequations}
\label{eq15}
\begin{eqnarray}
\Delta {\cal E}^{\perp} &\approx& -\, \frac{1}{24} \, \frac{\hbar
c}{D^3} \, \frac{{\omega_c}^2}{{\omega_p}^2} , \label{eq15a}\\
\Delta {\cal F}^{\perp} &\approx& -\, \frac{1}{8} \, \frac{\hbar
c}{D^4} \, \frac{{\omega_c}^2}{{\omega_p}^2}. \label{eq15b}
\end{eqnarray}
\end{subequations}
The third regime to be considered is the limit of short distances
($D \ll c/\omega_p$). Here one has to distinguish between two
regions: (i)~$\omega\leq k_\bot c \ll \omega_p$ and (ii)~$\omega_p
\ll \omega \leq k_\bot c$. In region~(i) the dielectric tensor
elements are given in Eqs.~(\ref{eq13}), so the reflection
coefficients $r_{sp}^{\perp}$, $r_{sp}^{\parallel}$ and $\Delta
r_{pp}$ are the same as those in the intermediate distance regime,
but we will now make an expansion for $r_{ss}$ and $r_{pp}$ around
$-1$ and $1$ respectively:
\begin{equation}
r_{ss}({\rm i}\omega , {\rm i}k_{\perp}) \approx -1 +
\frac{2k_{\perp}c}{\omega_{p}}, \ \ \ r_{pp}({\rm i}\omega , {\rm
i}k_{\perp}) \approx 1 - 2\frac{\omega^{2}}{\omega_{p}k_{\perp}c}
\label{eq16}
\end{equation}
In region~(ii) the dielectric tensor elements are given by:
\begin{subequations}
\label{eq17}
\begin{eqnarray}
\varepsilon_{xx}({\rm i}\omega) - 1 &\approx&
\frac{\omega_{p}^{2}}{\omega^{2}} \ll 1, \label{eq17a} \\
\varepsilon_{xy}({\rm i}\omega) &\approx&
\frac{\omega_{p}^{2}\omega_{c}}{\omega^{3}}. \label{eq17b}
\end{eqnarray}
\end{subequations}
We then find for region (ii):
\begin{subequations}
\label{eq18}
\begin{eqnarray}
r_{ss}({\rm i}\omega , {\rm i}k_{\perp}) &\approx& 0 , \
r_{pp}({\rm i}\omega , {\rm i}k_{\perp}) \approx
\frac{\omega_{p}^{2}}{2\omega^{2} + \omega_{p}^{2}}, \ \ \ \ \ \
\label{eq18a}
\\ r_{sp}^{\perp}({\rm i}\omega , {\rm i}k_{\perp}) &\approx&
-\frac{{\omega_p}^2\omega_c}{2} \, \frac{1}{(k_\perp c) (2\omega^2
+ {\omega_p}^2
)}, \label{eq18b} \\
r_{sp}^{\parallel}({\rm i}\omega , {\rm i}k_{\perp}) &\approx& -
\frac{\omega_{p}^{2}\omega_{c}}{2} \, \frac{\sqrt{\omega^{2} -
(k_{\perp}c)^{2}}}{(k_{\perp}c)^{2} (2\omega^{2} +
\omega_{p}^{2})}, \label{eq18c} \\
\Delta r_{pp}({\rm i}\omega , {\rm i}k_{\perp}) &\approx&
2\omega_{p}^{2}\omega_{c} \, \frac{\omega\sqrt{\omega^{2} -
(k_{\perp}c)^{2}}}{(k_{\perp}c)(2\omega^{2} +
\omega_{p}^{2})^{2}}. \label{eq18d}
\end{eqnarray}
\end{subequations}
With these approximations for the reflection coefficients in
regions (i) and (ii), one finds the following expressions for the
energies and forces (only the dominant term is given):
\begin{subequations}
\label{eq19}
\begin{eqnarray}
\Delta{\cal E}^{\perp} &\approx& - \frac{1}{16\sqrt{2}\pi}\,
\omega_{c}^{2}\sqrt{\omega_{p}}\,\frac{\hbar}{c^{3/2}D^{1/2}},
\label{eq19a} \\
\Delta{\cal F}^{\perp} &\approx& - \frac{1}{32\sqrt{2}\pi}\,
\omega_{c}^{2}\sqrt{\omega_{p}}\,\frac{\hbar}{c^{3/2}D^{3/2}},
\label{eq19b} \\
\Delta{\cal E}_{1}^{\parallel} &\approx& -
\frac{1}{96\sqrt{2}\pi}\,
\omega_{c}^{2}\sqrt{\omega_{p}}\,\frac{\hbar}{c^{3/2}D^{1/2}},
\label{eq19c} \\
\Delta{\cal F}_{1}^{\parallel} &\approx& -
\frac{1}{192\sqrt{2}\pi}\,
\omega_{c}^{2}\sqrt{\omega_{p}}\,\frac{\hbar}{c^{3/2}D^{3/2}},
\label{eq19d} \\
\Delta{\cal E}_{2}^{\parallel} &\approx&
\frac{1}{16\sqrt{2}\pi}\sum_{n=0}^{+\infty}\left[
\frac{(4n+3)!!}{(n+1)(4n+6)!!}\right]\frac{\omega_{c}^{2}}{\omega_{p}}
\frac{\hbar}{D^{2}}, \ \ \ \ \label{eq19e} \\
\Delta{\cal F}_{2}^{\parallel} &\approx&
\frac{1}{8\sqrt{2}\pi}\sum_{n=0}^{+\infty}\left[
\frac{(4n+3)!!}{(n+1)(4n+6)!!}\right]\frac{\omega_{c}^{2}}{\omega_{p}}
\frac{\hbar}{D^{3}}, \ \ \label{eq19f}.
\end{eqnarray}
\end{subequations}
From these expressions, it is obvious that $\Delta{\cal
E}^{\parallel} \approx \Delta{\cal E}_{2}^{\parallel}$ and
$\Delta{\cal F}^{\parallel} \approx \Delta{\cal
F}_{2}^{\parallel}$ for distances small enough. Note that for the
polar configuration, the exponent of the dependence with respect
to D obtained here in the short distance limit differs from the
one obtained in Ref.~[\onlinecite{PBru}]. This is due to the
effect of multiple reflections, which were neglected in
Ref.~[\onlinecite{PBru}] in this regime. It is interesting to note
that although the reflection coefficients are much smaller than 1
in this high-frequency limit, the effect of multiple reflections
is so important that the analytical dependence with D is modified.
This is a unique feature of the magnetic Casimir effect. \\
It is clear that in the polar configuration, the energies are
always negative. For the in-plane configuration however,
$\Delta{\cal E}_{2}^{\parallel}$ is positive, while $\Delta{\cal
E}_{1}^{\parallel}$ is negative, so the sign of the resulting
energy $\Delta{\cal E}^{\parallel} = \Delta{\cal
E}_{1}^{\parallel} + \Delta{\cal E}_{2}^{\parallel}$ will depend
on the magnitude of these two terms in the different regimes. As a
result, a change of sign of the interaction is observed; in the
long and intermediate distance regimes, the total energy
$\Delta{\cal E}^{\parallel}$ is negative, while for short
distances it is positive. So whether the magnetic Casimir
interaction is negative or positive depends on
the distance between the mirrors.\\
We numerically calculated Eqs.~(\ref{eq4}) and~(\ref{eq6}) and
compared them to the analytical expressions derived above. Details
of the numerical procedure will be given in
section~\ref{sectionnumeric}. The absolute values of the magnetic
Casimir forces per unit area (both numerical and analytical
results) for distances between 1 nm and 10 $\mu$m for the two
configurations are plotted in Fig.~\ref{fig1}. Since typically
$\tau \approx 10^{-13}$~s, the long distance regime ($D \gg c\tau
\approx 10 \ \mu$m) will not be visible in these plots. For the
plots, a Drude model is used with $\hbar \omega_{c} = 5.9$~meV and
$\hbar \omega_{p} = 9.85$~eV. As expected from the analytical
results, the force in the in-plane configuration will be larger
than that for the polar configuration for small enough distances
($D < 10$~nm). The discontinuity at $D \approx 40$~nm in the plot
of the in-plane case depicts the change of sign. The analytical
results are in pretty good agreement with the numerical
calculations.

\section{The Hybrid Model}

\begin{figure*}
\hspace*{-0.5cm}
\includegraphics{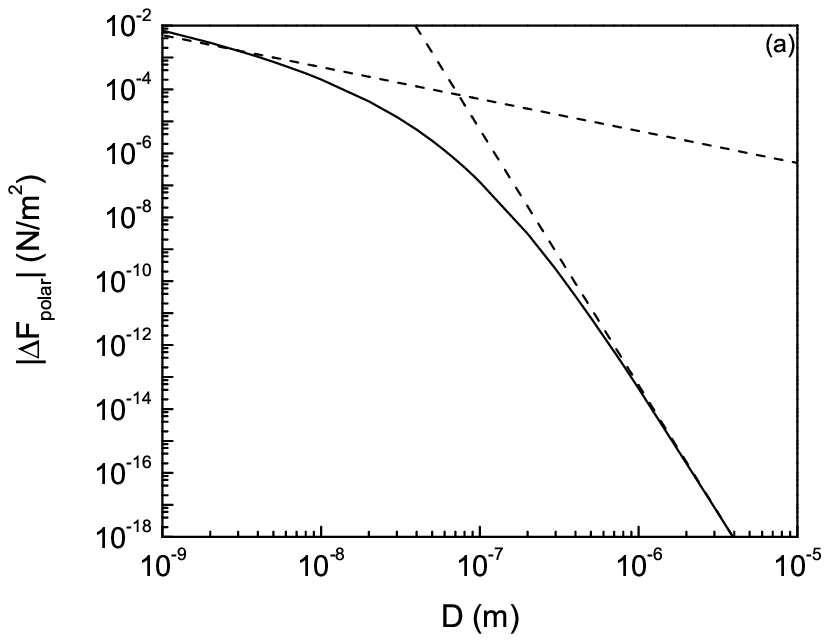}
\hspace*{-1cm}
\includegraphics{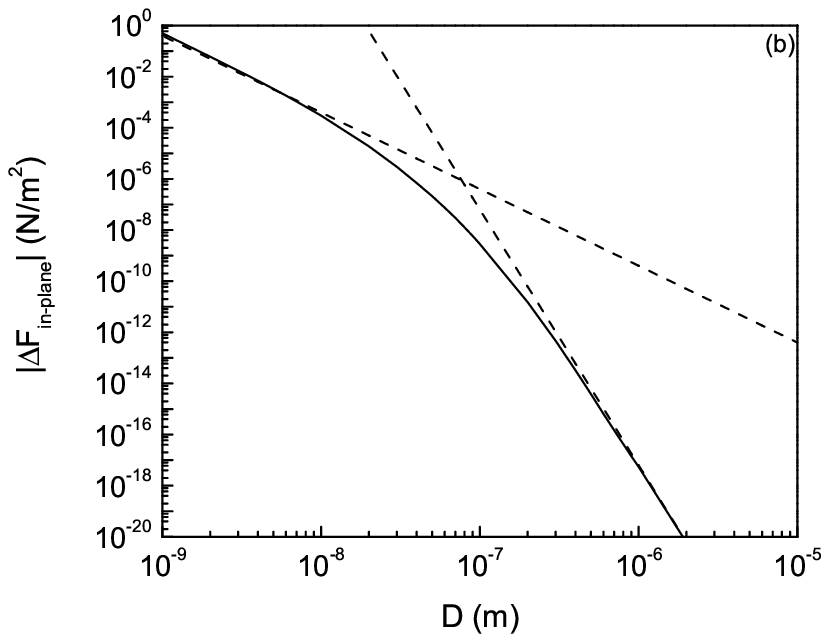}
\vspace*{-\baselineskip}

\hspace*{-0.5cm}
\includegraphics{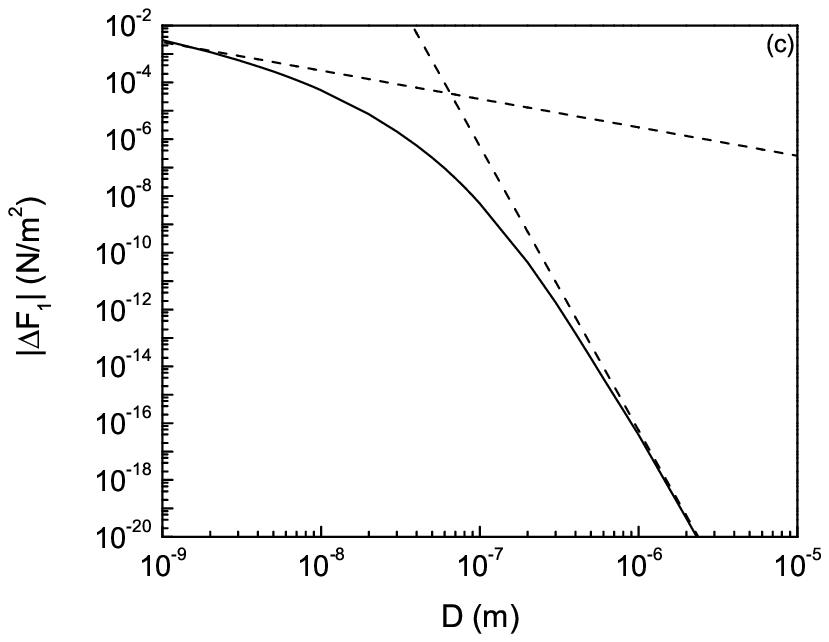}
\hspace*{-1cm}
\includegraphics{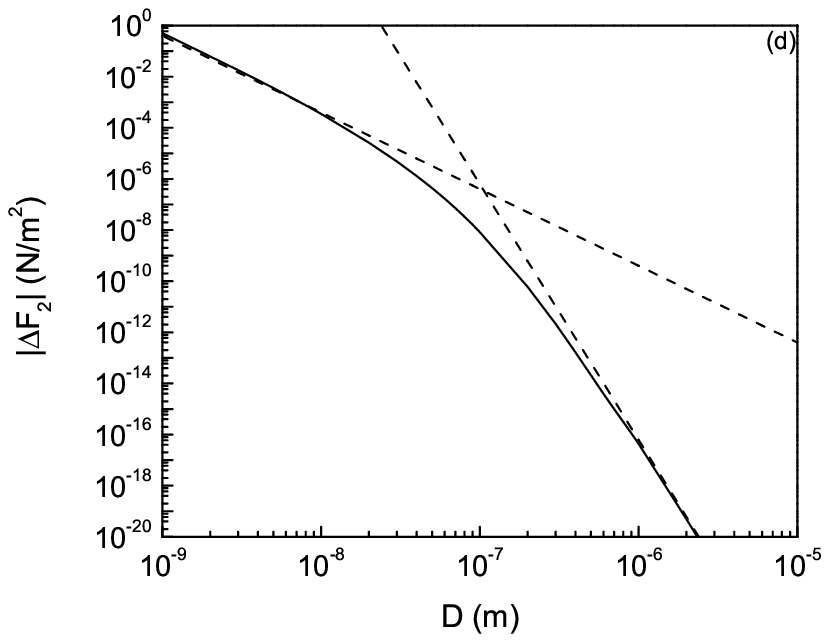}
\vspace*{-\baselineskip}
\caption{\label{fig2}Absolute values of
the magnetic Casimir force (per unit area) with the mirrors
described by the hybrid model. Numerical results (solid curves)
are compared with the analytical expressions (dashed curves) for
the polar configuration~(a), the in-plane configuration~(b), the
term $\Delta{\cal F}_{1}^{\parallel}$ resulting from the
longitudinal Kerr effect~(c) and the term from the transversal
Kerr effect $\Delta{\cal F}_{1}^{\parallel}$~(d).}
\end{figure*}

The Drude model is not very realistic. Although it describes
rather well the diagonal part of the dielectric tensor (except of
course for the effect of interband transitions, which are not very
important here), the off-diagonal part of the dielectric tensor is
poorly described. This is because the latter is dominated by
interband transitions. We therefore introduce a model (called
``hybrid model'') in which $\varepsilon_{xx}$ is described by a
plasma model,
\begin{equation}
\varepsilon_{xx}({\rm i}\omega) = 1 +
\frac{\omega_{p}^{2}}{\omega^{2}}, \label{eq20}
\end{equation}
and where $\varepsilon_{xy}$ is described by a single absorption
line interband transition:
\begin{equation}
\mbox{Re } \varepsilon_{xy} (\omega ) \approx \omega_0 \,
\varepsilon_{xy}^{\rm eff} \, \delta (\omega - \omega_0) .
\label{eq21}
\end{equation}
In real systems $\omega_{0}$ will be of the same order of
magnitude as $\omega_{p}$. The off-diagonal element of the
dielectric tensor at imaginary frequency can be obtained by the
following Kramers-Kr\"{o}nig relation:
\begin{equation}
\varepsilon_{xy} ({\rm i}\omega ) = \frac{2}{\omega\pi}
\int_0^{+\infty} \!\!\! {\rm d}\omega' \, \frac{{\omega'}^2 \,
\mbox{ Re }\varepsilon_{xy}(\omega' )}{{\omega'}^2 + \omega^2},
\label{eq22}
\end{equation}
and in this way we arrive at:
\begin{equation}
\varepsilon_{xy}({\rm i}\omega) = \frac{2}{\pi} \,
\frac{\omega_{0}^{3} \, \varepsilon_{xy}^{\rm
eff}}{\omega(\omega_{0}^{2} + \omega^{2})}. \label{eq23}
\end{equation}
For this model, we will only have two different integration
regimes; the long distance ($D \gg c/\omega_{p}$) and the short
distance regime ($D \ll c/\omega_{p}$). In the long distance
regime, the integrals in Eqs.~(\ref{eq4}) and~(\ref{eq6}) will be
dominated by the range $\omega \leq k_{\perp}c \ll \omega_{p}$. In
this range, we can approximate the dielectric tensor by:
\begin{subequations}
\label{eq24}
\begin{eqnarray}
\varepsilon_{xx}({\rm i}\omega) &\approx& \varepsilon_{xx}({\rm
i}\omega) - 1 \approx \frac{\omega_{p}^{2}}{\omega^{2}} \gg 1,
\label{eq24a} \\
\varepsilon_{xy}({\rm i}\omega) &\approx& \frac{2}{\pi} \,
\frac{\omega_{0} \, \varepsilon_{xy}^{\rm eff}}{\omega}.
\label{eq24b}
\end{eqnarray}
\end{subequations}
One then finds for the reflection coefficients:
\begin{subequations}
\label{eq25}
\begin{eqnarray}
r_{ss}({\rm i}\omega, {\rm i}k_{\perp}) &\approx& -1, \ \
r_{pp}({\rm i}\omega, {\rm i}k_{\perp}) \approx 1, \label{eq25a}
\\
r_{sp}^{\perp}({\rm i}\omega, {\rm i}k_{\perp}) &\approx& -
\frac{2}{\pi} \, \frac{\omega_{0} \, \varepsilon_{xy}^{\rm eff}}
{\omega_{p}^{3}}
\omega^{2}, \label{eq25b} \\
r_{sp}^{\parallel}({\rm i}\omega, {\rm i}k_{\perp}) &\approx& -
\frac{2}{\pi} \, \frac{\omega_{0} \, \varepsilon_{xy}^{\rm eff}}
{\omega_{p}^{4}} \sqrt{\omega^{2} - (k_{\perp}c)^{2}} \,
\omega^{2}, \label{eq25c} \\
\Delta r_{pp}({\rm i}\omega, {\rm i}k_{\perp}) &\approx&
\frac{4}{\pi} \, \frac{\omega_{0} \, \varepsilon_{xy}^{\rm eff}}
{\omega_{p}^{4}} \, \frac{\sqrt{\omega^{2} - (k_{\perp}c)^{2}} \,
\omega^{3}}{k_{\perp}c}. \label{eq25d}
\end{eqnarray}
\end{subequations}
With these approximations, we obtain for the magnetic Casimir
energies and forces (per unit area):
\begin{subequations}
\label{eq26}
\begin{eqnarray}
\Delta{\cal E}^{\perp} &\approx& -\frac{\pi^{2}}{210} \,
\frac{\omega_{0}^{2}(\varepsilon_{xy}^{\rm
eff})^{2}}{\omega_{p}^{6}} \, \frac{\hbar c^{5}}{D^{7}},
\label{eq26a} \\
\Delta{\cal F}^{\perp} &\approx& -\frac{\pi^{2}}{30} \,
\frac{\omega_{0}^{2}(\varepsilon_{xy}^{\rm
eff})^{2}}{\omega_{p}^{6}} \, \frac{\hbar c^{5}}{D^{8}},
\label{eq26b} \\
\Delta{\cal E}_{1}^{\parallel} &\approx& -\frac{\pi^{4}}{1050} \,
\frac{\omega_{0}^{2}(\varepsilon_{xy}^{\rm
eff})^{2}}{\omega_{p}^{8}} \, \frac{\hbar c^{7}}{D^{9}},
\label{eq26c} \\
\Delta{\cal F}_{1}^{\parallel} &\approx& -\frac{9\pi^{4}}{1050} \,
\frac{\omega_{0}^{2}(\varepsilon_{xy}^{\rm
eff})^{2}}{\omega_{p}^{8}} \, \frac{\hbar c^{7}}{D^{10}},
\label{eq26d} \\
\Delta{\cal E}_{2}^{\parallel} &\approx& \frac{\pi^{4}}{945} \,
\frac{\omega_{0}^{2}(\varepsilon_{xy}^{\rm
eff})^{2}}{\omega_{p}^{8}} \, \frac{\hbar c^{7}}{D^{9}},
\label{eq26e} \\
\Delta{\cal F}_{2}^{\parallel} &\approx& \frac{\pi^{4}}{105} \,
\frac{\omega_{0}^{2}(\varepsilon_{xy}^{\rm
eff})^{2}}{\omega_{p}^{8}} \, \frac{\hbar c^{7}}{D^{10}},
\label{eq26f} \\
\Delta{\cal E}^{\parallel} &\approx& \frac{\pi^{4}}{9450} \,
\frac{\omega_{0}^{2}(\varepsilon_{xy}^{\rm
eff})^{2}}{\omega_{p}^{8}} \, \frac{\hbar c^{7}}{D^{9}},
\label{eq26g} \\
\Delta{\cal F}^{\parallel} &\approx& \frac{\pi^{4}}{1050} \,
\frac{\omega_{0}^{2}(\varepsilon_{xy}^{\rm
eff})^{2}}{\omega_{p}^{8}} \, \frac{\hbar c^{7}}{D^{10}}.
\label{eq26h}
\end{eqnarray}
\end{subequations}
As in the Drude model, the force in the polar configuration will
be negative. However, the total force for the in-plane
configuration will be positive for the hybrid model in this distance regime.\\
In the limit of short distances, one has to distinguish between
three different integration ranges while performing the integrals
in Eqs.~(\ref{eq4}) and~(\ref{eq6}): region (i) where $\omega \leq
k_{\perp}c \ll \omega_{p}$, region (ii) $k_{\perp}c \gg
\omega_{p}, \ \omega \ll \omega_{p}$ and region (iii) $k_{\perp}c
\gg \omega_{p}, \ \omega \gg \omega_{p}$. In region (i), $r_{ss}$
and $r_{pp}$ are defined by Eq.~(\ref{eq16}), while
$r_{sp}^{\perp}, \ r_{sp}^{\parallel}$ and $\Delta r_{pp}$ are
given by Eqs.~(\ref{eq25b}-\ref{eq25d}). In regions (ii) and
(iii), we will do the calculations without multiple reflections
(i.e., we put $r_{ss}({\rm i}\omega, {\rm i}k_{\perp}) =
r_{pp}({\rm i}\omega, {\rm i}k_{\perp}) \approx 0$). In region
(ii), the dielectric tensor elements are given in
Eqs.~(\ref{eq24}), and the magneto-optical reflection coefficients
are given by:
\begin{subequations}
\begin{eqnarray*}
r_{sp}^{\perp}({\rm i}\omega, {\rm i}k_{\perp}) &\approx&
-\frac{1}{\pi} \, \omega_{0} \, \varepsilon_{xy}^{\rm eff} \,
\frac{\omega^{2}}{(k_{\perp}c)(2\omega^{2} + \omega_{p}^{2})},
\\
r_{sp}^{\parallel}({\rm i}\omega, {\rm i}k_{\perp}) &\approx&
-\frac{1}{\pi} \, \omega_{0} \, \varepsilon_{xy}^{\rm eff} \,
\frac{\sqrt{\omega^{2} - (k_{\perp}c)^{2}} \,
\omega^{2}}{(k_{\perp}c)^{2}(2\omega^{2} + \omega_{p}^{2})},
\\
\Delta r_{pp}({\rm i}\omega, {\rm i}k_{\perp}) &\approx&
\frac{4}{\pi} \, \omega_{0} \, \varepsilon_{xy}^{\rm eff} \,
\frac{\sqrt{\omega^{2} - (k_{\perp}c)^{2}} \,
\omega^{3}}{(k_{\perp}c)(2\omega^{2} + \omega_{p}^{2})^{2}}.
\end{eqnarray*}
\end{subequations}
In region (iii), the dielectric tensor can be approximated by:
\begin{subequations}
\label{eq27}
\begin{eqnarray}
\varepsilon_{xx}({\rm i}\omega) - 1 &\approx&
\frac{\omega_{p}^{2}}{\omega^{2}} \ll 1,
\label{eq27a} \\
\varepsilon_{xy}({\rm i}\omega) &\approx& \frac{2}{\pi} \,
\frac{\omega_{0}^{3} \, \varepsilon_{xy}^{\rm eff}}{\omega^{3}}.
\label{eq27b}
\end{eqnarray}
\end{subequations}
One then finds for the magneto-optical reflection coefficients in
region (iii):
\begin{subequations}
\label{eq28}
\begin{eqnarray*}
r_{sp}^{\perp}({\rm i}\omega, {\rm i}k_{\perp}) &\approx&
-\frac{1}{\pi} \, \omega_{0}^{3} \, \varepsilon_{xy}^{\rm eff} \,
\frac{1}{(k_{\perp}c)(2\omega^{2} + \omega_{p}^{2})},
\\
r_{sp}^{\parallel}({\rm i}\omega, {\rm i}k_{\perp}) &\approx&
-\frac{1}{\pi} \, \omega_{0}^{3} \, \varepsilon_{xy}^{\rm eff} \,
\frac{\sqrt{\omega^{2} -
(k_{\perp}c)^{2}}}{(k_{\perp}c)^{2}(2\omega^{2} +
\omega_{p}^{2})},
\\
\Delta r_{pp}({\rm i}\omega, {\rm i}k_{\perp}) &\approx&
\frac{4}{\pi} \, \omega_{0}^{3} \, \varepsilon_{xy}^{\rm eff} \,
\frac{\sqrt{\omega^{2} - (k_{\perp}c)^{2}} \,
\omega}{(k_{\perp}c)(2\omega^{2} + \omega_{p}^{2})^{2}}.
\end{eqnarray*}
\end{subequations}
With these equations by hand, we made an interpolation for the
reflection coefficients in region~(ii) and~(iii). This makes the
calculation of the integrals less labor-intensive because we can
calculate the two regions at the same time. We have put:
\begin{widetext}
\begin{subequations}
\label{eq29}
\begin{eqnarray}
r_{sp}^{\perp}({\rm i}\omega, {\rm i}k_{\perp}) &\approx&
-\frac{\omega_{0}^{3} \, \varepsilon_{xy}^{\rm eff}}{\pi} \,
\frac{\omega^{2}}{(k_{\perp}c)(2\omega^{2} +
\omega_{p}^{2})(\omega^{2} + \omega_{0}^{2})}, \label{eq29a}
\\
r_{sp}^{\parallel}({\rm i}\omega, {\rm i}k_{\perp}) &\approx&
-\frac{\omega_{0}^{3} \, \varepsilon_{xy}^{\rm eff}}{\pi} \,
\frac{\sqrt{\omega^{2} - k_{\perp}c^{2}} \, \omega^{2}
}{(k_{\perp}c)^{2}(2\omega^{2} + \omega_{p}^{2})(\omega^{2} +
\omega_{0}^{2})}, \label{eq29b}
\\
\Delta r_{pp}({\rm i}\omega, {\rm i}k_{\perp}) &\approx&
\frac{4\omega_{0}^{3} \, \varepsilon_{xy}^{\rm eff}}{\pi} \,
\frac{\sqrt{\omega^{2} - (k_{\perp}c)^{2}} \,
\omega^{3}}{(k_{\perp}c)(2\omega^{2} +
\omega_{p}^{2})^{2}(\omega^{2} + \omega_{0}^{2})}. \label{eq29c}
\end{eqnarray}
\end{subequations}
\end{widetext}
With these expressions for the reflection coefficients, we are
finally ready to calculate the magnetic Casimir energies and
forces for the short distance regime. The result is:
\begin{subequations}
\label{eq30}
\begin{eqnarray}
\Delta{\cal E}^{\perp} &\approx& -\frac{1}{4\sqrt{2}\pi^{3}} \,
\frac{\omega_{0}^{6} (\varepsilon_{xy}^{\rm
eff})^{2}}{(\omega_{p}+\sqrt{2}\omega_{0})^{3}} \,
\frac{\hbar}{c^{2}}\ln \left( \frac{c}{\omega^{\star} D} \right),
\ \ \ \ \
\label{eq30a} \\
\Delta{\cal F}^{\perp} &\approx& -\frac{1}{4\sqrt{2}\pi^{3}} \,
\frac{\omega_{0}^{6} (\varepsilon_{xy}^{\rm
eff})^{2}}{(\omega_{p}+\sqrt{2}\omega_{0})^{3}} \,
\frac{\hbar}{c^{2} D},
\label{eq30b} \\
\Delta{\cal E}_{1}^{\parallel} &\approx&
-\frac{1}{8\sqrt{2}\pi^{3}} \, \frac{\omega_{0}^{6}
(\varepsilon_{xy}^{\rm
eff})^{2}}{(\omega_{p}+\sqrt{2}\omega_{0})^{3}} \,
\frac{\hbar}{c^{2}}\ln \left( \frac{c}{\omega^{\star} D} \right),
\label{eq30c} \\
\Delta{\cal F}_{1}^{\parallel} &\approx&
-\frac{1}{8\sqrt{2}\pi^{3}} \, \frac{\omega_{0}^{6}
(\varepsilon_{xy}^{\rm
eff})^{2}}{(\omega_{p}+\sqrt{2}\omega_{0})^{3}} \,
\frac{\hbar}{c^{2} D},
\label{eq30d} \\
\Delta{\cal E}_{2}^{\parallel} &\approx& \frac{1}{64 \sqrt{2}
\pi^{3}} \, \frac{\omega_{0}^{6} (\varepsilon_{xy}^{\rm eff})^{2}
(\omega_{p} + 5 \sqrt{2} \omega_{0})}{\omega_{p} (\omega_{p} +
\sqrt{2} \omega_{0})^{5}} \,
\frac{\hbar}{D^{2}}, \label{eq30e} \\
\Delta{\cal F}_{2}^{\parallel} &\approx& \frac{1}{32 \sqrt{2}
\pi^{3}} \, \frac{\omega_{0}^{6} (\varepsilon_{xy}^{\rm eff})^{2}
(\omega_{p} + 5 \sqrt{2} \omega_{0})}{\omega_{p} (\omega_{p} +
\sqrt{2} \omega_{0})^{5}} \, \frac{\hbar}{D^{3}}, \ \ \
\label{eq30f}
\end{eqnarray}
\end{subequations}
with $\omega^{\star}$ a cut-off frequency of the order of the
plasma frequency $\omega_{p}$. It is clear that $\Delta{\cal
E}^{\parallel} \approx \Delta{\cal E}_{2}^{\parallel}$ and
$\Delta{\cal F}^{\parallel} \approx \Delta{\cal
F}_{2}^{\parallel}$ for distances small enough. The force for the
in-plane configuration is positive in this short distance regime
too, so there will not be a change of sign for this
model.\\
We did the same numerical calculations as for the Drude model and
compared the results with the analytical expressions in the
different regimes. The following parameters were used in our
hybrid model: $\hbar \omega_{p} = 9.85$~eV, $\hbar \omega_{0} =
3.9$~eV, $\varepsilon_{xy}^{\rm eff} = 1.5 \ 10^{-2}$ and we have
put $\omega^{\star} = 2 \, \exp(1) \, \omega_{p}$. In
Fig.~\ref{fig2} the numerical and analytical results for the
absolute value of the magnetic Casimir force are shown to be in
rather good agreement.

\section{Numerical Calculations on Fe} \label{sectionnumeric}

\begin{figure*}
\hspace*{-0.5cm}
\includegraphics{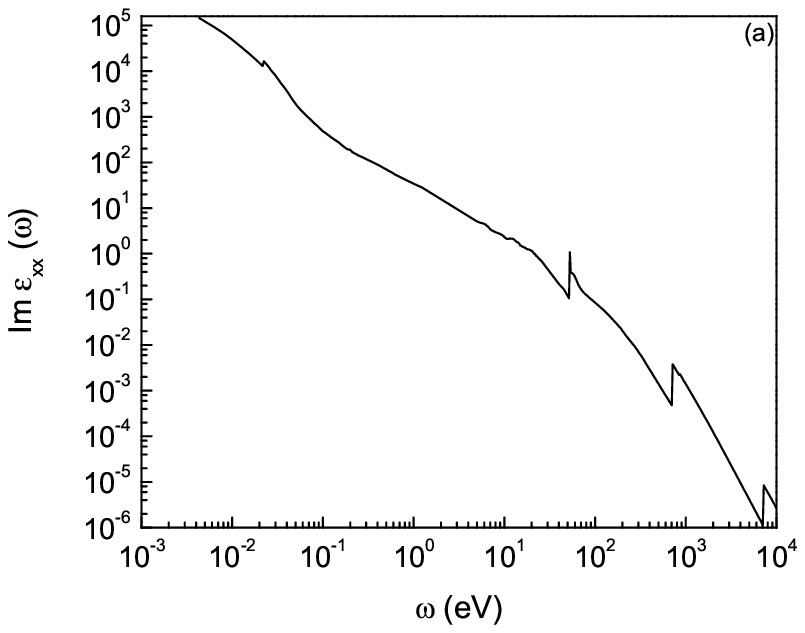}
\hspace*{-1cm}
\includegraphics{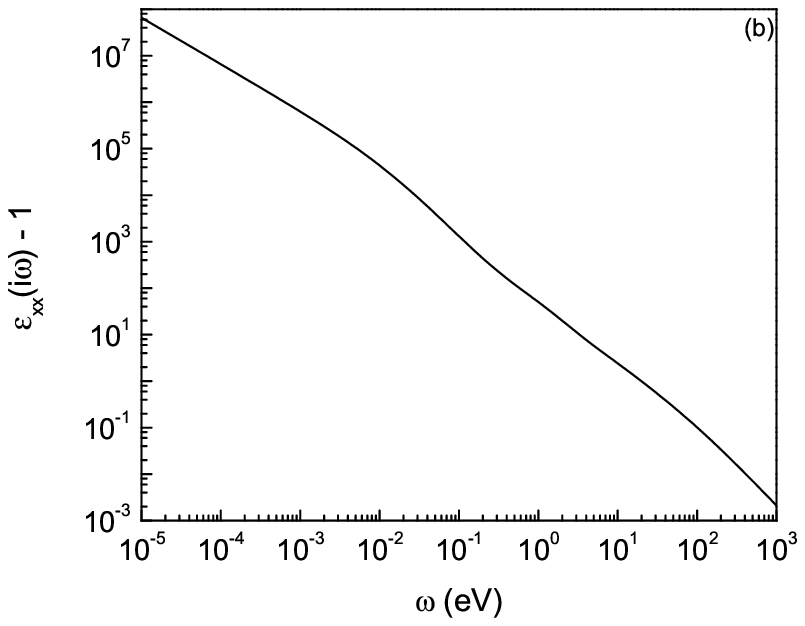}
\vspace*{-\baselineskip}
\caption{\label{fig3}The imaginary part
of the diagonal element of the dielectric tensor evaluated at real
frequencies~(a), and the diagonal element as a function of
imaginary frequency~(b).}
\end{figure*}

\begin{figure*}
\hspace*{-0.5cm}
\includegraphics{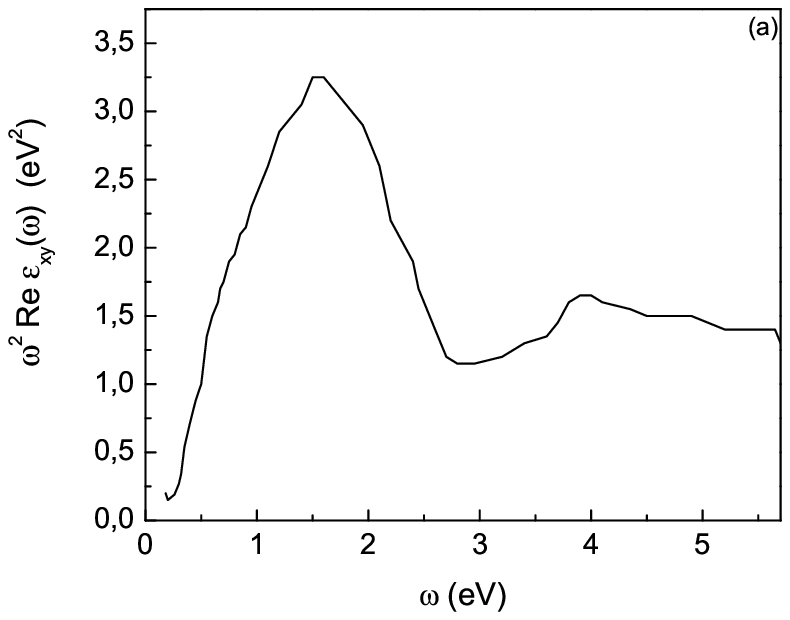}
\hspace*{-1cm}
\includegraphics{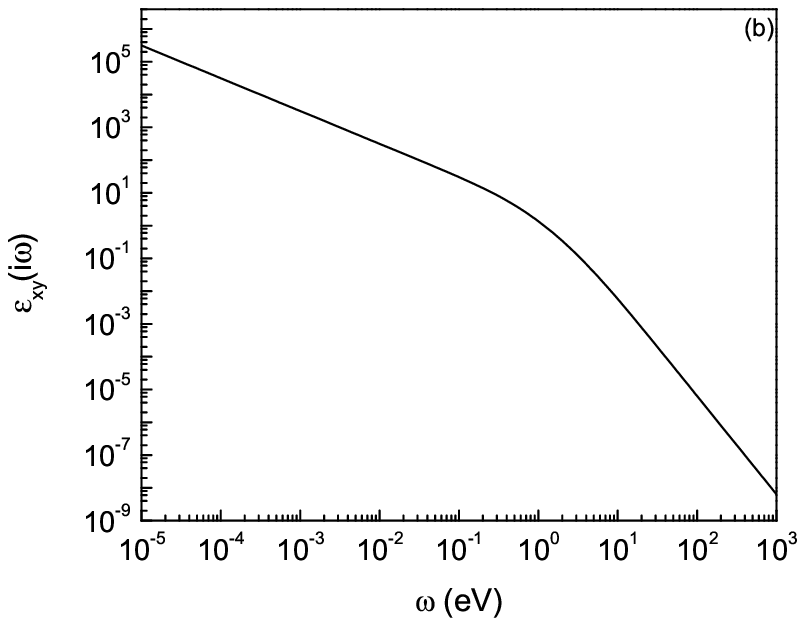}
\vspace*{-\baselineskip}
\caption{\label{fig4}The real part of
the off-diagonal element of the dielectric tensor evaluated at
real frequencies (multiplied by $\omega^{2}$)~(a), and the
off-diagonal element as a function of imaginary frequency~(b).}
\end{figure*}

The Drude and hybrid model will not provide an accurate
description for the dielectric tensor of the mirrors in a real
system. This is because interband transitions will start playing a
role at photon energies of a few~eV, and these are not contained
correctly in either one of the models. In order to obtain an
estimate of the magnitude of the magnetic Casimir force in such a
real system, it is necessary to perform numerical calculations of
Eqs.~(\ref{eq4}) and~(\ref{eq6}) where the reflection coefficients
are calculated with experimental data for the dielectric tensor.
In this section, we will present such calculations for a system
with iron plates. Similar calculations for the non-magnetic
Casimir force have already been
performed for Al, Au and Cu~\cite{SKLam,ALam}.\\
Experimental values for the imaginary part of
$\varepsilon_{xx}(\omega)$ for Fe are given
in~[\onlinecite{HandOpt}]. The diagonal element of the dielectric
tensor at imaginary frequency can then be obtained by the
causality relation:
\begin{equation}
\varepsilon_{xx} ({\rm i}\omega ) = 1+\frac{2}{\pi}
\int_0^{+\infty} \!\!\! {\rm d}\omega' \, \frac{\omega' \, \mbox{
Im }\varepsilon_{xx}(\omega' )}{{\omega'}^2 + \omega^2}.
\label{eq31}
\end{equation}
Of course it is impossible to perform the numerical integration
over the entire interval $[0,+\infty]$, so we have to define our
integration range in more detail. In our calculations, the
complete range of data extending from 4~meV to 10~keV available
in~[\onlinecite{HandOpt}] was used, along with a Drude model below
4~meV as shown in Fig.~\ref{fig3}. The following parameters for
the Drude model were found by extrapolation of the available data
at low frequencies: $\hbar \omega_{p} = 3.5$~eV and $\hbar/\tau =
19$~meV. The quantity $\varepsilon_{xx}({\rm i}\omega) -1$
calculated in this way is shown in Fig.~\ref{fig3} to decay
roughly as $\omega^{-3/2}$ (for $\omega > \hbar/\tau$), so it can
not be completely described by a Drude (or plasma) model.

\begin{figure*}
\hspace*{-0.5cm}
\includegraphics{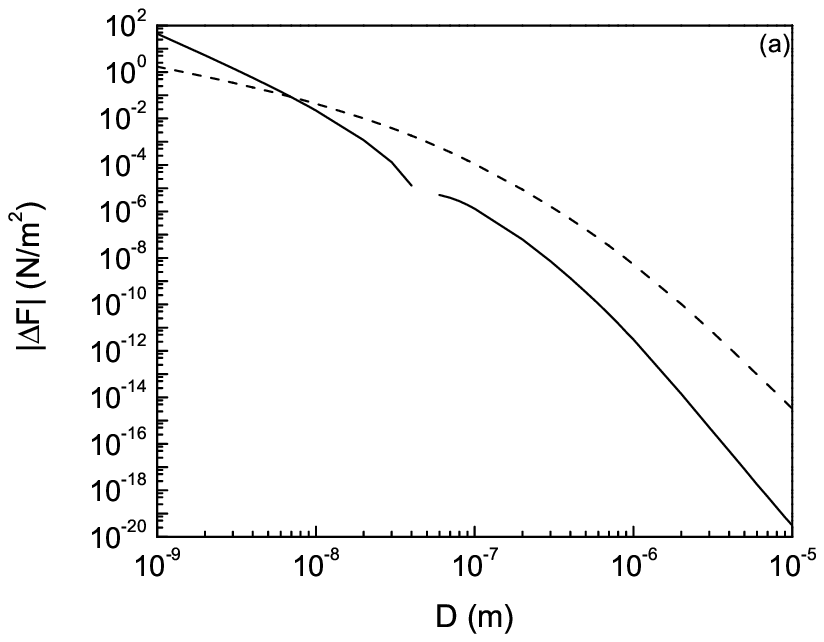}
\hspace*{-1cm}
\includegraphics{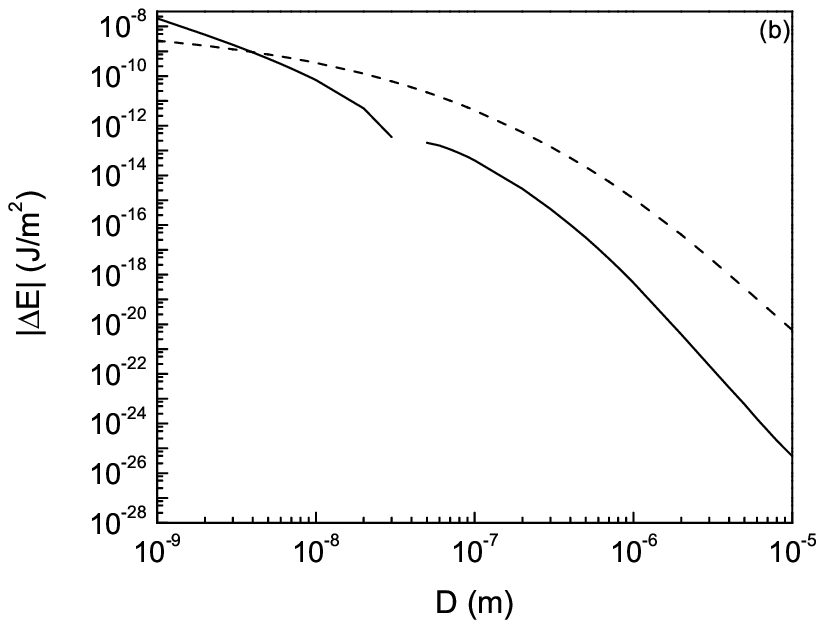}
\vspace*{-\baselineskip}
\caption{\label{fig5}Absolute values of
the magnetic Casimir force per unit area~(a) and the magnetic
Casimir energy per unit area~(b) between two iron plates (of
infinite lateral extension). The solid curve corresponds to the
in-plane configuration, while the dashed curve describes the polar
configuration.}
\end{figure*}

Experimental data for the off-diagonal element of the dielectric
tensor is rather scarce. Some data for $\mbox{Re }
\varepsilon_{xy}(\omega)$ can be found in~[\onlinecite{Land}].
They are shown in Fig.~\ref{fig4}. With the causality
relation~(\ref{eq22}) it is then possible to calculate
$\varepsilon_{xy}({\rm i}\omega)$. Since we only have data
available between 0.1~eV and 6~eV, we had to perform the
integration in Eq.~(\ref{eq22}) over this range. This, of course,
is a rather rough approximation. The results of the calculation
depicted in Fig.~\ref{fig4} show that $\varepsilon_{xy}({\rm
i}\omega)$ decays the same way as in our hybrid model (cf.
Eqs.~(\ref{eq24b}) and~(\ref{eq27b})).

The magnetic Casimir force and energy are now calculated by
numerical integration of Eqs.~(\ref{eq4}) and~(\ref{eq6}). We are
interested in plate separations between 1~nm and 10~$\mu$m. These
separations correspond to frequencies in the range
$10^{-2}$~-~$10^{2}$~eV, so we will have to perform the
integration between say $10^{-5}$~eV and $10^{4}$~eV.
Fig.~\ref{fig5} shows the resulting force and energy (per unit
area) for the polar and in-plane configuration. In the short
distance limit, the force decays as $D^{-2}$ for the polar
configuration and as $D^{-3}$ for the in-plane case. For long
distances we find a $D^{-6}$ power law for the polar configuration
and $D^{-8}$ for the situation with magnetization parallel to the
plates. A change of sign of the interaction for the in-plane
configuration is also visible from the figure (the discontinuity
at $D = 50$~nm).  The power laws differ (except for the in-plane
configuration at short distances) from those obtained for the
Drude and hybrid model. This can be explained as due to the
different behavior (because of interband transitions) of the
dielectric tensor for Fe compared to that of the models. In view
of future experimental investigations of the effect, distances
$D>10$~nm are the most interesting. In this range, the effect will
be greatest for the polar configuration. For two parallel plates
of Fe (with infinite lateral extension) the force per unit area in
this configuration is approximately 40~mN/m$^{2}$ at $D=10$~nm,
and decays to 0.1~mN/m$^{2}$ at $D=100$~nm. Whether such forces
can be observed experimentally will be discussed in the next
section.

\section{Experimental Setup}
Since it is hard to experimentally maintain two parallel plates
uniformly separated by distances smaller than a micron, one of the
plates is most often replaced by a lens-shaped mirror. Recently, a
number of experiments has been performed using this geometry to
measure the non-magnetic Casimir force with an
AFM~\cite{UMoh,ARoy,BWHar}. For this plate-cylinder geometry, the
Casimir force can be obtained from the plate-plate geometry by
means of the force proximity theorem~\cite{JBlo}:
\begin{equation}
\Delta F = 2\pi R \, \Delta {\cal E}(D). \label{eq32}
\end{equation}
In this equation, $R$ is the radius of curvature of the
lens-shaped mirror, and $\Delta {\cal E}(D)$ is the Casimir energy
per unit area for the configuration with two plates. One has to be
careful to distinguish between $\Delta F$ and $\Delta {\cal F}$;
the former is the force for the plate-lens geometry, while the
latter is a force per unit area for two parallel plates. With the
numerical results from the previous section, we are able to
estimate the magnitude of the magnetic Casimir force in the
plate-lens geometry for Fe. If we take $R=100 \mu$m and a distance
$D = 50$~nm, a force $|\Delta F | \approx 10$~fN is found for the
polar configuration. In the in-plane configuration, the force will
be two orders of magnitude smaller. Such small forces can probably
not be measured with the AFM technique. However, sensitivities of
0.1 to 10~fN in ``magnetic resonant force microscopy (MRFM)'' have
been reported~\cite{DRug,JASid}. More recently, the detection of
forces in the attonewton range has been achieved~\cite{TDStow,
PMoh}.

A possible MRFM setup is already discussed in detail in
Ref.~[\onlinecite{PBru}]. A thin film ($\approx$~10~nm) of
ferromagnetic material with hard magnetization is deposited on a
substrate which is placed on a piezo-electric actuator. The
lens-shaped mirror is attached to a cantilever by first depositing
a small droplet of polymer on the cantilever, which can then be
covered by evaporation with a thin ($\approx$~10~nm) layer of soft
ferromagnet (such as permalloy). In this way, one is able to
create the lens-shape, with a curvature radius of say $100 \
\mu$m. The distance between the samples can be controlled easily
with the piezo-actuator. By applying an ac magnetic field, one is
able to modulate the magnetization of the soft sample at the
resonance frequency of the cantilever. This will generate an
oscillating magnetic Casimir force that causes the cantilever to
vibrate. The deflection of the cantilever can then be measured
with a laser. In this way, the magnetic force ($\Delta F = F_{AF}
- F_{FM}$) can be measured. The force resolution achievable using
a freely vibrating cantilever is fundamentally limited by
intrinsic thermomechanical noise. This force noise can be
controlled by the geometry of the cantilever; one needs a high Q
cantilever that is thin, narrow and long to obtain the best
sensitivity. With ultrathin silicon cantilevers, force resolutions
in the attonewton range have been obtained~\cite{TDStow}. More
information on the sensitivity of MRFM can be found
in~[\onlinecite{TDStow,JASid}].

Since the non-magnetic Casimir effect is independent of the
magnetization direction of the samples, only the magnetic
contribution to the Casimir effect will be measured by using this
modulation technique. Parasitic electrostatic forces (caused by a
difference in potential between the magnetic samples) are also
taken care of automatically in this way. The exchange interaction
between the samples does not contribute much at the separations
of interest ($D > 10$~nm). Another parasitic magnetostatic
interaction is the dipole interaction between the ferromagnets.
This dipole force can be made as small as needed by taking a
ferromagnetic plate with sufficiently large lateral extension and
sufficiently small thickness. The plate should also be as
uniformly magnetized as possible. With a plate of radius 1~cm,
and a thickness of 10~nm, this parasitic magnetostatic force can
be estimated to be below 1~attoNewton. Interaction of the soft
sample with the ac magnetic field will yield a signal at two
times the modulation frequency, so this can be filtered out
effectively by using a lock-in amplifier. Thus with MRFM, it
should be possible to measure the magnetic Casimir interaction
without much influence of other effects.

\section{Conclusion}
In this paper, the magnetic Casimir interaction discovered
in~[\onlinecite{PBru}] was generalized to the case where the
magnetization is parallel to the plates. The calculations for the
Drude model in the short-distance limit were revised, and another
model was introduced. The behavior of the interaction was
discussed in the different distance regimes, and it is seen that
the interaction in the two models decays quite different with
interplate distance. Numerical calculations for a real system with
iron plates were also presented. Here we used experimental data
for the dielectric tensor of the mirrors. The results from this
numerical work on Fe could not be fitted by one of the introduced
models, because interband transitions play a prominent role in Fe,
and these were not implemented correctly in
the models. \\
It was made acceptable that the new Casimir magnetic interaction
can be measured with magnetic resonance force microscopy. However,
to obtain an accurate comparison of the theory with eventual
experimental results, more work is needed on the theoretical side.
A detailed analysis of the geometrical effects would be valuable;
the proximity force theorem does not provide reliable estimations
at a level of accuracy of a few percent. Also one has to consider
other corrections already calculated for the non-magnetic Casimir
effect; e.g. surface roughness corrections would probably play an
important role~\cite{ARoy2}. Finally, more experimental data on
the off-diagonal element of the dielectric tensor for several
ferromagnetic materials is also necessary in order to obtain a
better estimate of the magnitude of the interaction from the
numerical procedure presented.

\end{document}